# Two-dimensional single-crystal photonic scintillator for enhanced X-ray imaging

Tatsunori Shibuya[1,2,*], Eichi Terasawa[1,2,*], Hiromi Kimura[1,2], and Takeshi Fujiwara[1,2,*]

[1] Non-Destructive Imaging Research Team, Integrated Research Center for Resilient Infrastructure, National Institute of Advanced Industrial Science and Technology (AIST), 1-1-1 Umezono, Tsukuba 305-8568, Japan.

[2] Research Institute for Measurement and Analytical Instrumentation (RIMA), National Metrology Institute of Japan (NMIJ), National Institute of Advanced Industrial Science and Technology (AIST), 1-1-1 Umezono, Tsukuba 305-8568, Japan.

* Corresponding authors:
Tatsunori Shibuya: t-shibuya@aist.go.jp
Eichi Terasawa: wisdom.terasawa@aist.go.jp
Takeshi Fujiwara: fujiwara-t@aist.go.jp

**Abstract**

The evolution of X-ray detection technology has significantly enhanced sensitivity and spatial resolution in non-destructive imaging of internal structure. However, the problem of low luminescence and transparency of scintillator materials restricts imaging with lower radiation doses and thicker materials. Here, we propose a two-dimensional photonic scintillator for single crystal and demonstrate that the optical guiding effect emerging from the structure reduces luminescence leakage and increases the signal intensity by around a factor of 2 from 200 to 450 kV. This approach has the potential to enhance the output rate by an order of magnitude. The photonic structure features a fine array pitch and large-scale detection area with fast fabrication time. Our scheme paves the way for high sensitivity X-ray imaging.

**Introduction**

X-ray imaging offers a distinct advantage for probing thick materials owing to its high penetration depth, enabling non-destructive visualization of internal structures inaccessible to optical methods **[1,2,3,4,5]**. Recent advances in X-ray detection technologies have enabled high-spatiotemporal resolution and sensitivity imaging, facilitating detailed information retrieval at lower radiation doses for the human body **[6,7,8,9,10,11,12]**, lighter elements and thicker packaging measurements for batteries **[13,14,15,16,17,18,19]** or semiconductor devices **[20,21,22,23,24,25]**, time-resolved in-situ measurements in internal combustion engines or motors [26,27,28], high-speed and high-resolution baggage inspection **[29,30,31,32]**, welding void inspection **[33,34,35,36]**, and material deterioration diagnosis **[37,38]** applications. While the ability to penetrate materials and internal defects offers substantial advantages, conventional detection schemes suffer from spatial—temporal integration and low sensitivity **[39,40,41]**, particularly under higher-energy X-ray regimes (>100 keV) or thick material conditions (>10 mm). These problematic detection schemes obscure localized defect differences, limiting access to deeper insights into the material. To overcome these challenges and fully realize the potential of advanced X-ray detection systems, a highly sensitive scintillator is critically needed to efficiently convert high-energy X-ray signals into detectable visible light with minimal loss.

X-ray detectors that convert X-ray photons into visible light influence imaging quality and radiation dose **[42].** X-ray detectors are classified as either direct or indirect converters. Direct converters offer a high spatial resolution but relatively low sensitivity **[43,44]**, whereas indirect converters can provide high sensitivity but low spatial resolution **[45,46]**. For this reason, the employment of an indirect converter is preferred over that of a direct converter at the high-energy X-ray regime, which enhances the signal intensity deriving from the X-ray absorption efficiency. In the case of indirect convertors, sheet-type scintillators (for example, GOS: $Gd_2O_2S$:Tb **[47]** and CsI:Tl **[48,49]**) are used to achieve both high resolution and a wide field of view. In the mid-level X-ray energy regime (from 20 to 100 kV) for indirect converter scintillators, several previous researchers implemented a powdered CsI scintillator by packing the powder into RIE-etched bulkhead arrays of Si or $SiO_2$ and sintering them; the light scattering from the powder hinders the performance of the partitioning **[50,51,52]**. For the higher energy X-ray regime (from 200 to 500 kV), ceramic/polymer scintillators exhibit high X-ray absorption but become increasingly opaque to luminescence as their thickness increases **[42,53,54,55]**. In the case of a single-crystal scintillator, transparency is related to minimal internal light scattering, and increasing the crystal thickness can enhance luminescence and optical output. However, the diffusion of emitted luminescence into the bulk leads to image smearing, degrading both brightness and spatial resolution. Specifically, scintillator performance is determined by (1) luminescence intensity and centre wavelength for high quantum efficiency, (2) density and thickness for high X-ray absorption, and (3) transmission efficiency. To fully enhance detection performance, single-crystal scintillators with low crystal defect density and high dopant concentration are among the most effective options **[56,57]**.

Here, we propose a two-dimensional photonic scintillator for a single crystal. By comparing two X-ray imaging studies using the photonic structure crystal (Fig.1, left) and an untreated homogeneous crystal (Fig.1, right), we demonstrate that the optical guiding effect emerging from the photonic structure reduces luminescence leakage and has high brightness and resolution.

**Results**

**Concept of two-dimensional single-crystal photonic scintillator**

Assuming $\theta_c < \pi/4$, the luminescence extraction efficiency $T$ is described as follows.

$$T = \frac{P_d}{P_t} = \frac{P_d}{P_d + P_s + P_u + P_r} \tag{1}$$

Here, $P_t$ is total luminescence power and it is constant, $P_d$ is the light power towards the detector, $P_s$ and $P_u$ are the light power escaping from the side and upper plane, respectively, and $P_r$ is the confined light into the bulk caused by internal total reflection. This photonics scintillator works under the two schemes: Fresnel reflection mode and diffuse reflection mode. In Fresnel reflection mode, From the viewpoint of Fresnel's equation and internal multiple reflection, it can be shown that the relationship between the power passing through the detector plane $P_d$, the angle of incidence $\theta$, the critical angle $\theta_c$, reflection rate of the front side $R_1(\theta)$ and back side $R_2(\theta)$ satisfies the following formula:

$$P_d = \frac{P_0}{2}\left(\int_0^{\theta_c} \sin\theta\, d\theta \frac{R_1(\theta)(1-R_2(\theta))}{1-R_1(\theta)R_2(\theta)} + \int_{\pi-\theta_c}^{\pi} \sin\theta\, d\theta \frac{1-R_2(\theta)}{1-R_1(\theta)R_2(\theta)}\right) \tag{2}$$

From equation (2), on the condition that side wall and bottom/top wall are perpendicular to each other, $\theta_c$ is independent of lengths (mainly, scintillator thickness $l$ and pixel size $d$) at arbitrary points (in case of $\tan\theta < d/l$, $P_d$ is independent, and in case of $\tan\theta > d/l$, since the symmetry is preserved by the side wall, $P_d$ becomes changeless), that is to say, the geometry of this photonic structure and output power to detector behaves independently.

In diffuse reflection mode, since the luminescence is diffuse-reflected on the reflection boundary, the $P_r$ is changed to $P_{diff}$ and the part of $P_r$ became a detectable $P_{diff}{'}$ and, $P_d$, $P_s$ and $P_u$ are changed to $P_d{'}$, $P_s{'}$ and $P_u{'}$ due to the surface diffuse reflection, therefore the luminescence extraction efficiency of diffuse reflection $T_d$ is described as follows.

$$T_d = \frac{P_d{'} + P_{diff}{'}}{P_d{'} + P_s{'} + P_u{'} + P_{diff}} \tag{3}$$

Since the diffuse reflection breaks the conservation of $\theta_c = $ const., the geometry depends on the output power.

As a promising candidate for high quality pixelization, femtosecond laser material processing **[58]** can provide ultrahigh precision, as the electron density achieves the critical plasma density much faster than the heat diffusion into bulk. As the required laser intensity is $10^{12}$ to $10^{13}$

W/cm² at most, the parameter range is relatively tolerant, and the technique is commonly used to control spatial pulse shaping. One such technique is Bessel beam optics, which generates a spatial profile extended in the depth direction for percussioning **[59,60]** and dicing **[61]**. The Bessel beam optics technique has been widely applied to the processing of amorphous glass, providing diameters from 100 nm to 10 μm and an aspect ratio up to 1,000:1.

**Experimental proof of optical guiding in photonic scintillator**

According to ray-tracing simulations of the photonic scintillator (Fig. 2 **A** to **F**), from equation 2, since Fresnel reflection mode only has the function for light confinement effect and reflection mode increase the total amount of light from equation 3, Fresnel and diffuse reflection mode operate as high resolution and high brightness.

The X-ray luminescence intensity was measured and compared with the side direction (Fig.2 **H**) and transmittance (Fig.2 **I**) arrangements to verify the optical guiding effect of luminescence. The measurement used X-rays with a tube voltage of 80 kV and no stainless-steel filter between the source and the detector. The luminescence from the photonic scintillator observed from the side was clearly reduced. By contrast, in the transmittance arrangement, it is increased compared to that from the homogeneous scintillator. This result confirms that the presence of the partitioning suppresses light leakage and induces an optical guiding effect. This observation indicates that the brightness enhancement of the photonic structure shown in Fig. 4 results from suppressed light leakage. This photonic structure intrinsically works on the same principle as optical fibers, so the significant refractive index difference between the cavities created by laser irradiation and homogeneous GAGG: $n_{GAGG} - n_0 = 0.9$ is responsible for the optical guiding effect, which can be explained by Fresnel reflection.

This two-dimensional photonic scintillator can be readily adopted as a practical replacement for thicker scintillators, since it can block most of the X-rays (60%@100keV) at a thickness of 800 μm. It provides twice the brightness over a wide energy range. This light-enhancement mechanism is in principle valid for thicker scintillators, but technical constraints of the laser treatment limit it to a thickness of 2 mm. Nevertheless, a scintillator thickness of 2 mm covers almost all applications of X-rays.

**Brightness of two-dimensional photonic scintillator**

As shown in Fig. 3 **A**, by placing a 15 mm stainless steel plate right behind the X-ray source, most of the X-ray spectrum below 100 keV could be cut off, and a quasi-monochromatic high-energy X-ray beam was extracted. The count rates of images were compared for each accelerating voltage by simply subtracting the dark current images (X-ray off mode) from the X-ray image (X-ray on mode). The images in Fig. 3 **B** were acquired simultaneously, and the distance between the source and the detector was sufficiently large compared to the detector size, allowing the spatial distribution of X-rays to be approximated as uniform. The image acquisition times for 100, 200, 300, 400, and 450 kV are 5, 3, 2, 1, and 1 s, respectively, and the data were normalized to acquisition time. The signal intensity of the photonic scintillator was increased by 2.57, 1.99, 2.01, and 2.02 times at 200, 300, 400, and 450 kV, respectively. The photonic structure was more effective at relatively low X-ray energies yet still exhibited approximately twice the brightness even at the higher energy of 450 kV (Fig. 3 **C**).

In previous studies, photonic scintillators have not been shown to operate above 200 kV. On the other hand, if we focus only on the light enhancement factor, the present scintillator is inferior to that of Ref. **[62]**. However, since the photonic structuring approach by Bessel beam is a process inside the bulk, it can coexist with most surface photonic structuring, and the combined light enhancement factor is enhanced to 12 to 15.

**Image contrast evaluation for a thick stainless steel using photonic scintillator**

Figure 4 shows the experimental setup (Fig. 4 **A**) and the image contrast evaluation of the visualizing step edge sample (Fig. 4 **B**) with a thickness difference of 1/16 inch (1.59 mm) viewed through a stainless-steel plate with a thickness of 70 mm. The results of Fig. 4 **C**, d are shown normalized to transmission values, in which X-rays pass through the position where only GAGG is present, to a maximum value of 1.0. At a plate thickness of 70 mm, X-rays below 150 kV are nearly eliminated, making it suitable for evaluating the sensitivity to high-energy X-rays. Even under these conditions, a difference between 1/16 inch and 1/8 inch (1.59 mm) could be distinguished. As the stainless-steel plate is 70 mm thick, the transmittance of X-rays below 150 kV is less than 1%, making it suitable for evaluating the sensitivity of high-energy X-ray over 200 kV. Even under these conditions, the photonic scintillator distinguished between 1/16 inch and 1/8 inch (1.59 mm), corresponding to a thickness difference of 2.2%.

**Evaluation of femtosecond laser material processing**

Previously, dicing of crystal scintillators was challenging owing to the formation of microcracks or cleavage during fabrication. Nevertheless, after extensive cutting and polishing efforts, only a limited array of structures, such as 25×25 pixels with a pixel size of 400 μm, was reported **[56]**. However, fabricating smaller pixels, such as 200 μm, and arrays exceeding 100×100 pixels, had remained challenging. A rapid fabrication process was needed without compromising spatial–temporal resolution and luminescence intensity. As a result, a pixel dicing rate of 210 mW s/mm was achieved, enabling faster fabrication of smaller features than the conventional method. For example, dicing a 100 μm box with a laser scanning speed of 380 mm/s on a 4-inch square using an 80-watt femtosecond laser is estimated to take 270 s, which is a realistically achievable time. Bessel beam optics are applicable to configurations with a width margin of 10–20 μm and an aspect ratio ranging from 100:1 to 1,000:1 **[63]**. As Bessel beam optics allows for magnitude reduction optics, when the material is relatively thin, such as 10 to 100 μm, reducing the typical margin width to 100 nm scale is possible. Therefore, the net pixel size could be reduced to approximately 1–100 μm, allowing for higher pixel density. Additionally, applying laser treatment enables fabrication of photonic structures up to a height of 2 mm. With the minimum pixel size reduced to 100 μm, the spatial resolution may be enhanced to 10 lines pairs per millimeter (lp mm$^{-1}$). Furthermore, combining with other photonic fabrication techniques on the surface to suppress light scattering at the surface may also be possible **[64,65,66]**. There's a possibility that employing stealth dicing **[67]** assisted by phosphoric acid **[68]** leads to reduce the pixel size of GAGG. The scalability of laser material processing approach provides the truly industrial large-scale photonic scintillator with 300 mm×300 mm detector.

**DISCUSSION**

We have demonstrated a two-dimensional single-crystal photonic scintillator with the following essential features: (1) The relationship between the luminescence light extraction efficiency and the geometric configuration of the scintillator was elucidated, enabling photonic control through both diffuse and Fresnel reflection modes. (2) By inscribing the photonic structure inside of millimeter-thick scale scintillator, a highly efficient light absorption rate and light enhancement factor across a broad X-ray energy range was realized. (3) Submillimeter-scale pixel dimensions and centimeter-scale structural patterns were rapidly fabricated via femtosecond laser material processing. (4) We demonstrated the potential to surpass other light enhancement methods by combining our approach with additional surface photonic techniques.

Owing to the 31.8° critical angle for internal reflection, up to 17.5% of the optical output could be coupled to the detector directly without Fresnel reflection. In addition, when a large flat surface is coated with a diffuse reflector (Teflon, reflectivity > 90%), the optical output rate could reach up to 64.7% theoretically, about three times higher than that of a conventional scintillator, even without accounting for detection efficiency [48,49]. Furthermore, it is expected that, depending on the material thickness, X-ray detectable efficiency of this photonic scintillator increases and finally achieves over one order of magnitude higher than homogeneous GAGG, which has the same imaging resolution.

  In this study, a method for fabricating a two-dimensional single-crystal photonic scintillator using femtosecond laser material processing with Bessel beam optics was implemented. The photonic structure formed rapidly and exhibited sufficient mechanical strength. The Bessel beam optics enabled the single crystal scintillator to achieve pixel sizes four times smaller, with processing times significantly shorter than those of conventional manual methods. The wall height achieved by Bessel beam dicing was 0.7 mm, comparable to or greater than that obtained via the powder bed method using reactive ion etching. Comparison of X-ray images from the laser-treated photonic and untreated homogeneous scintillators showed that the photonic structure has at least twice the brightness in the high-energy X-ray regime. This fabrication scheme is expected to be applicable to single-crystal photonic scintillator technologies for X-ray and gamma-ray detectors.

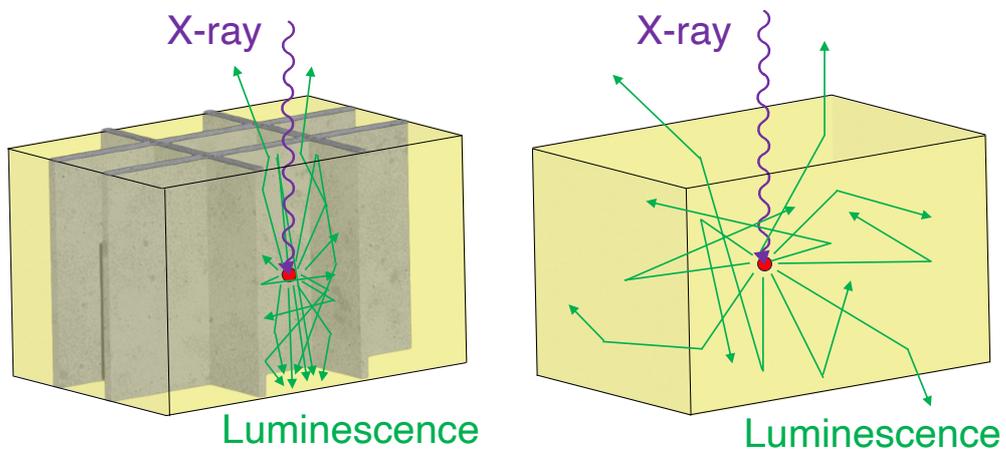

**Fig. 1 Photonic structures prevent light diffusion.** Schematic of luminescence process for two-dimensional photonic scintillator: photonic structure (left) and homogeneity (right).

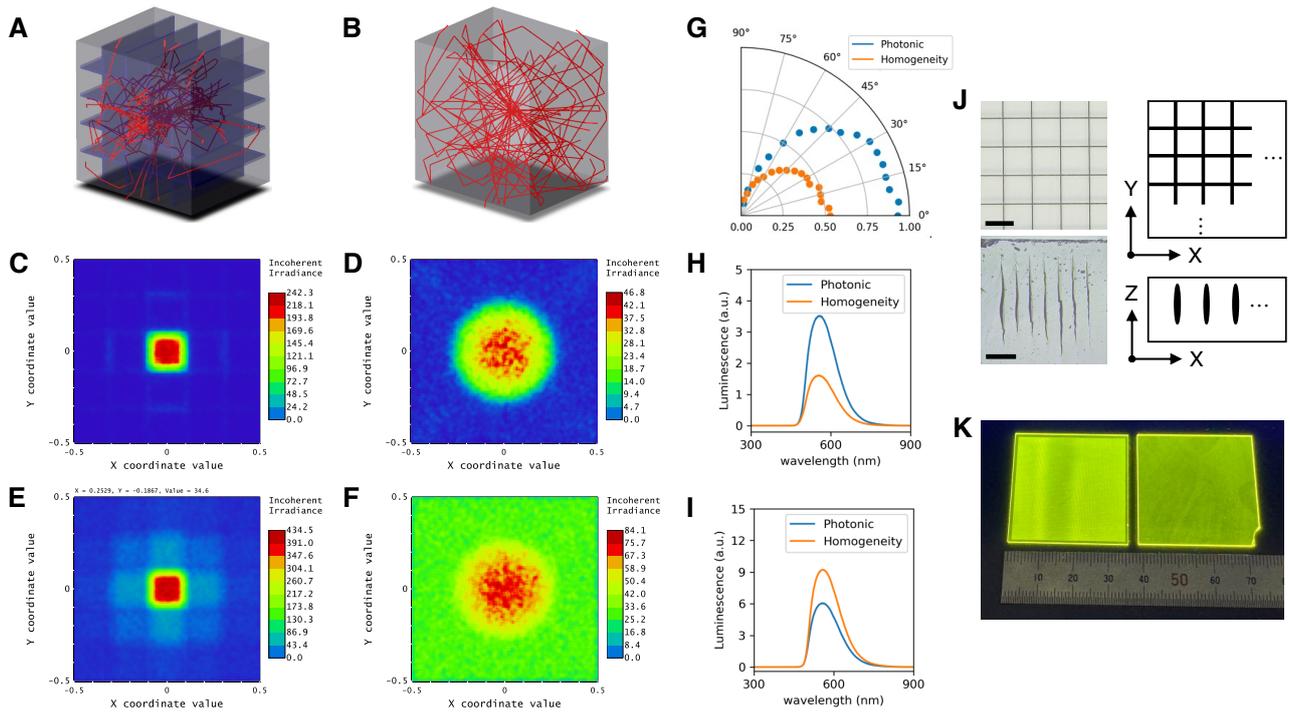

**Fig. 2 Experimental demonstration of two-dimensional photonic scintillator.** (**A** to **F**) Ray tracing simulation of (**A**) photonic model and (**B**) homogeneous model. Intensity distribution of (**C** and **E**) photonic and (**D** and **F**) homogeneity. The top surface is defined as (**C** and **D**) Fresnel reflection and (**E** and **F**) diffuse reflection. (**G** to **I**) Angular dependence of X-ray luminescence intensity. (**G**) Angular dependence intensity comparison following X-ray excitation at 80 keV accelerating voltage of photonic (blue) and homogeneity (orange) structure, having the same material size. The emission is partially blocked by the mask with the pinhole diameter of 2 mm and the detector was placed 50 mm away from the pinhole. (**H**) Detection direction perpendicular to the line through X-ray source point and the center point of scintillator. (**I**) Detection direction parallel to the line through X-ray source point and center point of scintillator. (**J**) surface image (top) and cross-sectional image (bottom). (**K**) Picture of, left, photonic structure and, right, homogeneity structure made of gadolinium aluminum gallium garnet ($Gd_3Al_2Ga_3O_{12}$:Ce, GAGG:Ce).

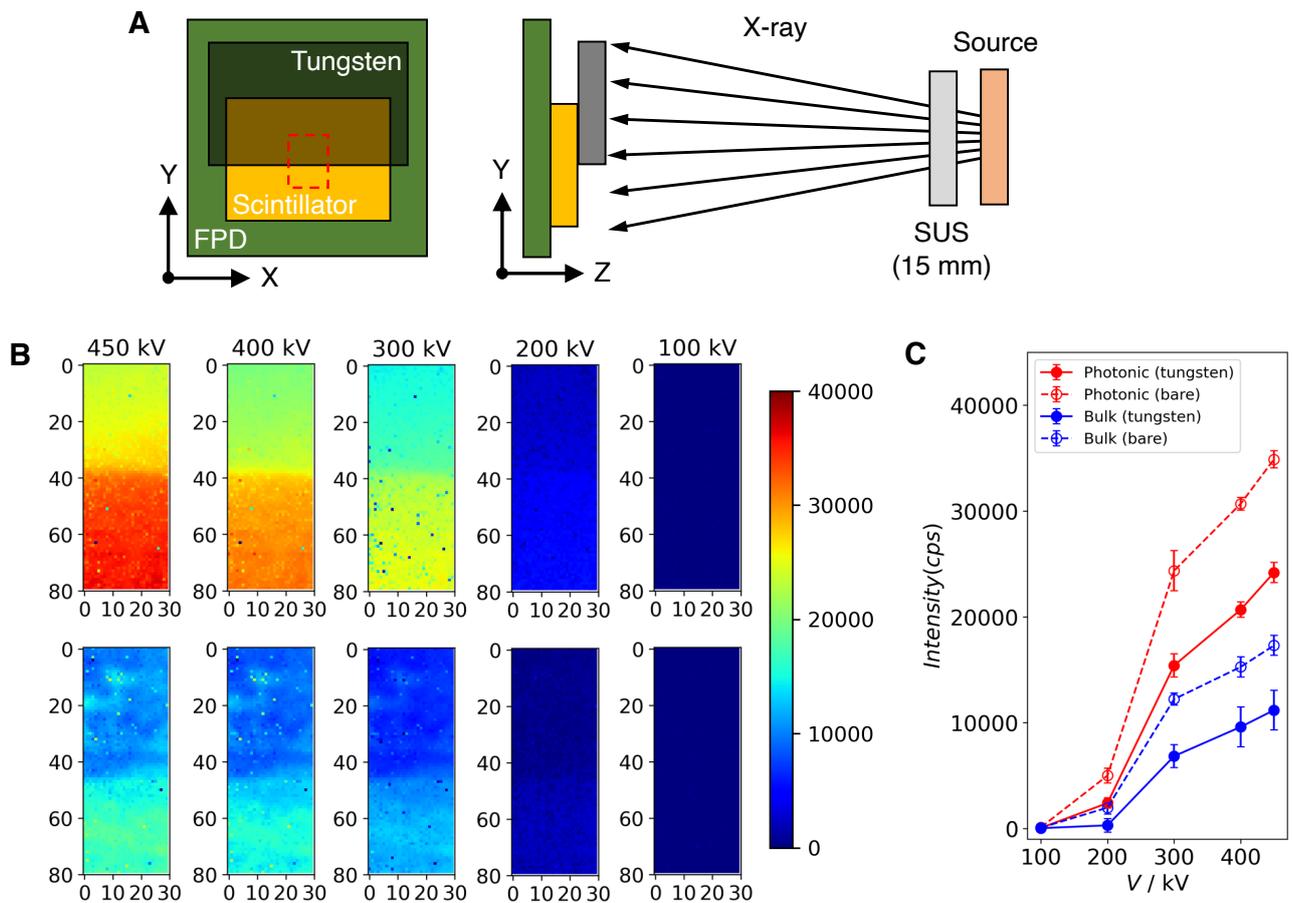

**Fig. 3 Brightness comparison of two-dimensional photonic and homogeneous scintillators at each X-ray energy.** (**A**) Experimental setup of high-energy X-ray imaging. (**B** and **C**) Luminescence comparison of the photonic scintillator GAGG (top) and the homogeneous scintillator GAGG (bottom) at energy regime from 100 kV to 450 kV. This part corresponds to the red dashed line frame in (**A**). The image acquisition times for 100, 200, 300, 400, and 450 kV are 5, 3, 2, 1, and 1 s, respectively, and the data are normalized by the acquisition time.

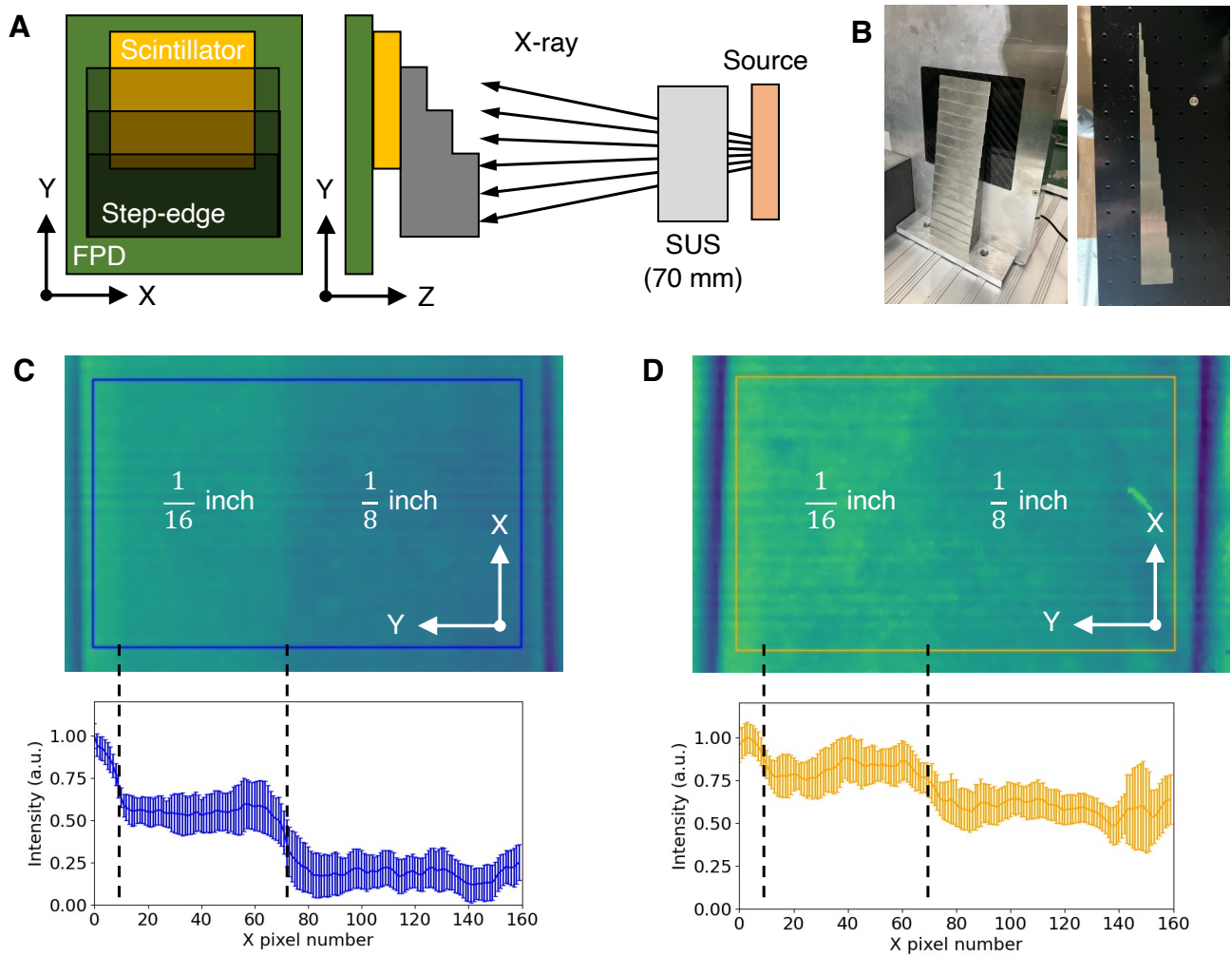

**Fig. 4 High-energy X-ray imaging of step edge sample viewed through stainless steel with a thickness of 70 mm at X-ray tube voltage of 450 kV.** (**A**) Experimental setup of high-energy X-ray imaging. The X-ray exposure time is 3 s. (**B**) Picture of the step edge sample. (**C**) Results of two-dimensional photonic scintillator. X-ray image (top) and line intensity difference of 1/16 inch and 1/8 inch with averaging of blue box in X direction (bottom). (**D**) Results of homogeneous scintillator. X-ray image (top) and line intensity difference of 1/16 inch and 1/8 inch with averaging of yellow box in X direction (bottom).

## MATERIALS AND METHODS

### Performance of homogeneous GAGG: Ce

The GAGG: Ce crystal exhibits (1) superior X-ray absorption rate owing to high density (6.6 g cm$^{-3}$) and the presence of high-Z element (gadolinium), (2) high transparency rate at 520 nm (Ce$^{3+}$ 5d$_1$→4f emission) close to the maximum quantum efficiency of silicon-based photodiodes, (3) high luminescence intensity of 46,000 photons/MeV [69,70], (4) less hygroscopic without sealing, and (5) a fast decay time of 100 ns [71,72]. The refractive index at 540 nm is 1.9, and the critical angle for internal reflection is 31.8°. According to ref. [73], the optical output rate was approximately 7% for air coupling. Recently, large crystals have become manufacturable, and research is underway into the dicing process for gamma-ray detectors.

### Femtosecond laser material processing with Bessel beam optics

For photonic structure fabrication, a Ti:sapphire laser with a center wavelength of 800 nm (Coherent Inc., Libra), pulse duration of 70 fs, maximum pulse energy of 3 mJ and repetition rate of 1 kHz [74] was used. A schematic of the partitioning process and physical samples is shown in Fig. S4 a, b and c. The typical transmission wavelength of GAGG spans 500–800 nm, and the optical energy is absorbed via multiphoton absorption. The laser power was adjusted by a combination of a λ/2 waveplate and a polarizer, and during laser irradiation, the pulse energy was 63.1±2.3 μJ per pulse. The sample was mounted on an XYZ automatic stage, which has a constant scan speed of 300 μm/s, and the optical axis was set perpendicular to the sample surface. The Bessel-beam optics consisted of an axicon lens with a physical angle of 2.0° (Thorlabs Inc.), plano–convex lens with a focal length of 200 mm (Thorlabs Inc.) and an objective lens with a magnification of 20 (Mitsutoyo Co., Mplan Apo NIR 20X). The optics optimization was conducted using "reZonator" software. A schematic of the laser irradiation setup, ray-tracing simulation of Bessel optics, and fabricated GAGG crystal are shown in Fig. 2 a. Note that for the XZ-plane cross section used in setting conditions, the pitch scale was 100 μm. Beam alignment was confirmed by observing the propagation of the characteristic ring-shaped Bessel beam in the far field. The Z-position of irradiation was explored to be a position where no nonlinear energy absorption occurred at the surface, and the bottom of the beam was determined to be 50 μm from the surface. The scribe structure on the surface was confirmed as a scratch mark caused by the lower surface damage threshold and is formed only on the top-surface layer. Ultimately, a two-dimensional square lattice structure with a pitch of 200 μm, maximum array height of 720 μm, and total size of 33 mm, i.e., 165 pixels, was fabricated within the single-crystal photonic scintillator (Fig. S4 b and c).

### X-ray induced luminescence and imaging experiments

X-ray induced luminescence experiment was done using X-ray source (XMS-803) built by Kinki Roentgen Industrial CO., LTD., and the luminescence was capture using spectrometer (QEPro, Ocean Photonics) and a photomultiplier (H10682-110, Hamamatsu Photonics). For X-ray imaging experiment, MesoFocus 450 from Comet AG was employed as an X-ray source. The source size is typically 63 μm, and broadband X-ray with maximum voltage of 450 kV is generated. X-rays pass through a filter made from stainless steel with a thickness of 15 mm or 70 mm, and image a tungsten sample with a thickness of 1 mm placed close to the detector. The distance between the source and the detector was 1000 mm. The visible light from the scintillator was converted into electrical signals by an indium gallium zinc oxide flat panel detector with a pixel size of 200 μm, which corresponds one-to-one to the pixel positions, and was read out as image signals.